\documentclass[%
 reprint,
superscriptaddress,
 amsmath,amssymb,
 aps,pre,
floatfix,
]{revtex4-1}

\usepackage{graphicx}
\usepackage{dcolumn}
\usepackage{bm}
\usepackage{mathtools}
\usepackage{color}
\usepackage{xspace}

\newcommand{\BUDGET}{\omega}
\newcommand{\dC}{d_{\text{c}}}
\newcommand{\Dc}{D_{\text{c}}}
\newcommand{\xC}{x_{\text{c}}}
\newcommand{\kC}{k_{\text{c}}}
\newcommand{\kCF}{k_{\text{cf}}}
\newcommand{\dGCF}{\Delta G_{\text{cf}}}
\newcommand{\kM}{k_{\text{m}}}
\newcommand{\xM}{x_{\text{m}}}
\newcommand{\dM}{d_{\text{m}}}
\newcommand{\Dm}{D_{\text{m}}}

\begin{document}

\preprint{APS/123-QED}

\title{Pulling cargo increases the precision of molecular motor progress}

\author{Aidan I. Brown}
\email{aibrown@ucsd.edu}
\affiliation{Department of Physics, Simon Fraser University, Burnaby, BC, V5A1S6 Canada}
\affiliation{Department of Physics, University of California, San Diego, San Diego, California 92093}

\author{David A. Sivak}
\email{dsivak@sfu.ca}
\affiliation{Department of Physics, Simon Fraser University, Burnaby, BC, V5A1S6 Canada}

\begin{abstract}
Biomolecular motors use free energy to drive a variety of cellular tasks, including the transport of cargo, such as vesicles and organelles. We find that the widely-used `constant-force' approximation for the effect of cargo on motor dynamics leads to a much larger variance of motor step number compared to explicitly modeling diffusive cargo, suggesting the constant-force approximation may be misapplied in some cases. We also find that, with cargo, motor progress is significantly more precise than suggested by a recent result.
For cargo with a low relative diffusivity, the dynamics of continuous cargo motion---rather than discrete motor steps---dominate, leading to a new, more permissive
bound on the precision of motor progress which is independent of the number of stages per motor cycle.
\end{abstract}

\maketitle

\section{Introduction}

Biomolecular machines, driven by free energy consumption, perform a variety of cellular roles~\cite{alberts98}. Motors, such as kinesin, myosin, and dynein tow cargoes along subcellular filaments for rapid and directed transport~\cite{vale03}. The stochastic dynamics of these motors have been studied for decades~\cite{hackney96}, with recent experimental work observing their operation with improving resolution~\cite{isojima16}.

An important aspect of motor dynamics is the response to external forces, as the primary role of many \emph{in vivo} motors is to tow cargoes, which impose drag forces. External forces are frequently modeled as constant across all motor cycles and cycle stages~\cite{fisher99,fisher01,thomas01,qian04,lau07,hinczewski13,qian16,wagoner16,hwang17}, recapitulating sophisticated experiments which use feedback to maintain near-constant resisting forces on the forward motion of molecular motors~\cite{visscher99,clemen05,clancy11,nicholas15}.

Molecular motors are also excellent systems in which to investigate statistical fluctuations at the nanoscale. Fluctuations in molecular motor progress have been related to the 
number of stages~\cite{svoboda94,fisher01,koza02,koza02b} and energy dissipation budget~\cite{barato15,barato15b} for each motor cycle.

In this work we investigate simple theoretical models of molecular motors towing cargo, with cargo represented either implicitly as a constant force or explicitly with diffusive dynamics. We find that constant-force models for cargo do not in general reproduce the step-number distributions of explicit diffusive-cargo models. The step-number Fano factor for diffusive cargo is also substantially less than when predicted solely from motor characteristics~\cite{svoboda94,fisher01,koza02,koza02b,barato15,barato15b}. We demonstrate that diffusing cargo introduces a large number of effective stages to each motor cycle, allowing motors to increase the precision of their progress by pulling cargo, and often rendering the number of motor states irrelevant to the precision of motor progress. Our new, more permissive bound on the precision of motor progress, expressed in Eq.~\eqref{eq:fanocontinuous}, only depends on the energy dissipation budget per cycle.

\section{Model}

\begin{figure}[tb]
	\centering
	\hspace{-0.0in}
	\begin{tabular}{c}
		\hspace{-0.200in}\includegraphics[width=3.4in]{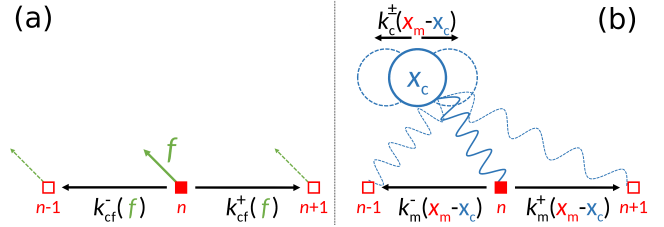}
	\end{tabular}
	\caption{\label{fig:Diagrams} 
		{\bf Constant-force and diffusing-cargo models.} (a) Constant-force model. Motor at site $n$ (filled red square) and position $\xM$ can transition forward to site $n+1$ at rate $\kCF^+$ or reverse to site $n-1$ at rate $\kCF^-$. The rates $\kCF^{\pm}$ are functions of a constant opposing force $f$. (b) Diffusing-cargo model. The motor and the cargo (blue circle) are linked, so that forward and reverse motor and cargo transition rates $\kM^{\pm}$ and $\kC^\pm$, respectively, are a function of motor position $\xM$ and cargo position $\xC$, which vary with cargo and motor motion.
	}
\end{figure}

Molecular motor dynamics are commonly described as stochastic transitions between discrete states. For a simple one-state unicyclic model, every forward cycle results in a forward step of the motor and consumption of the free energy budget $\BUDGET$ per cycle. For `constant-force' motor dynamics (shown schematically in Fig.~\ref{fig:Diagrams}a), we use static transition rate constants
\begin{equation}
\label{eq:cfrateconstants}
\kCF^+ = \kCF^0 e^{\BUDGET-\dGCF} \ \ \text{and} \ \ \kCF^- = \kCF^0 \ ,
\end{equation}
with $\kCF^+$ and $\kCF^-$ the rate constants of forward and reverse transitions, respectively. $\kCF^0$ is a `bare' rate constant setting the transition timescale, chosen to set $\kCF^0=1$. $\dGCF$ is a constant free-energy difference representing the effect of a load (\emph{e.g.}~towing cargo) on rate constants. $\dGCF$ is frequently set to $f\dM$, with $f$ the constant opposing force and $\dM$ the motor step size. We set energies to units of $k_{\text{B}}T$, for Boltzmann's constant $k_{\text{B}}$ and temperature $T$. 
The resulting motor diffusivity is~\cite{hwang17}
\begin{equation}
\label{eq:diffusivity}
D_{\text{cf}} = \frac{1}{2}\left(\kCF^+ + \kCF^-\right)\dM^2 \ .
\end{equation}

We also consider a similar `diffusing-cargo' kinetic model of a motor taking forward and reverse steps while coupled by a Hookean spring (with spring constant $k$) to a cargo also taking discrete steps (Fig.~\ref{fig:Diagrams}b). The motor has transition rate constants
\begin{equation}
\label{eq:ecrateconstants}
\kM^{+} = \kM^{0} e^{\BUDGET - \Delta G^+_{\text{sm}}(\xM-\xC)} \ \ \text{and} \ \ \kM^{-} = \kM^{0} \ ,
\end{equation}
and the cargo has transition rate constants
\begin{equation}
\label{eq:cargorateconstants}
\kC^{\pm}=
\begin{cases}
\kC^{0}, &\text{if }\Delta G_{\text{sc}}^{\pm}(\xM-\xC) \leq 0\\
\kC^{0}e^{-\Delta G_{\text{sc}}^{\pm}(\xM-\xC)}, &\text{if }\Delta G_{\text{sc}}^{\pm}(\xM-\xC) > 0 \ ,
\end{cases}
\end{equation}
providing standard response to an applied force (see Supplementary Material).
$x_{\text{m}}$ is the motor position and $x_{\text{c}}$ is the cargo position. $\kM^{0}$ and $\kC^{0}$ are the bare rate constants for the motor and cargo, respectively; we choose a timescale such that $\kM^0=1$. $G_{\text{s}}=\tfrac{1}{2}k(\xM - \xC)^2$ is the free energy of the spring linking motor to cargo. $\Delta G_{\text{sm}}^{+} = k\dM(\xM - \xC) + \tfrac{1}{2}k\dM^2$ is the change in $G_{\text{s}}$ due to a forward motor step of size $\dM$, while $\Delta G_{\text{sc}}^{\pm} = \mp k\dC(\xM - \xC) + \tfrac{1}{2}k\dC^2$ is the change in $G_{\text{s}}$ due to a forward ($+$) or reverse ($-$) cargo step of size $\dC$.
Cargo diffusivity in the absence of a motor is $\Dc = \kC^0(\dC)^2$, with $\dC$ held fixed in all simulations and $\kC^{0}$ varied to adjust $\Dc$. $\Dm = \tfrac{1}{2}\kM^0 \dM^2(e^{\BUDGET}+1)$ is the diffusivity of a motor without cargo.

The Fano factor for net number $n$ of motor steps taken can be related to 
the
number of stages in a motor cycle $N$ 
by 
$\sigma_{n}^2/\mu_{n} \geq 1/N$, where $\sigma^2_{n}$ is the variance of the net forward cycles and $\mu_{n}$ is the mean number of net forward cycles~\cite{svoboda94,fisher01,koza02,koza02b}. Recent work by Barato and Seifert~\cite{barato15,barato15b} predicts a `thermodynamic uncertainty relation,' which further tightens the previous $N$-dependent bound by incorporating the free energy dissipation budget $\BUDGET$ per cycle:
\begin{equation}
\label{eq:bsfano}
\text{Fano factor} \equiv \frac{\sigma^2_{n}}{\mu_n} \geq \frac{1}{N}\coth{\frac{\BUDGET}{2N}} \ ,
\end{equation}
This bound is for constant-affinity dynamics, and has received substantial attention~\cite{hwang17b,gingrich16,polettini16,gingrich17,proesmans17,sartori15,nguyen16,ouldridge17,hwang18}.
For a motor cycle with a single step ($N=1$), Eq.~\eqref{eq:bsfano} becomes an equality, $\sigma^2_n/\mu_n = \coth\tfrac{\BUDGET}{2}$.

Our results primarily use relatively small values for the cargo diffusivity, consistent with the low diffusivity of \emph{in vivo} motor cargoes such as vesicles and organelles. A diffusivity of $\sim$$10^{-3}\mu\text{m}^2/\text{s}$ has been measured for 80nm-radius beads in the cytosol~\cite{luby-phelps00}, suggesting vesicles and organelles have diffusivities $\Dc \sim 10^{-5}-10^{-3}\mu\text{m}^2/\text{s}$. Using experimental kinesin rate constants~\cite{vu16} to estimate $\Dm$ suggests $\Dc/\Dm \sim 10^{-3}-10^{-1}$ (see Supplementary Material).

\section{Results}

\begin{figure}[tb]
	\centering
	\hspace{-0.0in}
	\begin{tabular}{c}
		\hspace{-0.200in}\includegraphics[width=3.4in]{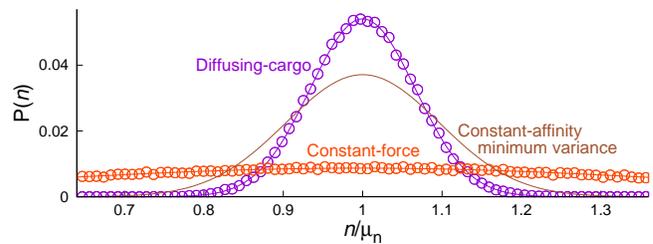}
	\end{tabular}
	\caption{\label{fig:PositionDists}
		{\bf Anomalously narrow step-number distribution.} Step-number probability distribution $P(n)$ as a function of normalized step number $n/\mu_{n}$. Diffusing-cargo model step-number distribution is narrower than constant-force model and narrower than expected from minimum Fano factor for a constant-affinity cycle. Step-number distribution from $10^5$ samples of numerical diffusing-cargo model (purple circles) after nondimensionalized time $t^{\ast} \equiv \kM^0 t = 10^3$ with $\BUDGET=4$, diffusivity ratio $\Dc/\Dm = 10^{-3}$, spring constant $k^* \equiv k \dM^2 = 1$, and relative cargo step size $\dC/\dM = 0.001$, along with Gaussian fit (purple curve). Orange circles: Step-number distribution from $10^5$ samples of numerical constant-force model with $\BUDGET=4$ and equal mean step number $\mu_{n}$ through choice of $\dGCF = 3.8945$. Orange curve: Gaussian distribution predicted by mean $\mu_n$ and Eq.~\eqref{eq:diffusivity}. Brown curve: Gaussian distribution with mean $\mu_{n}$ and the constant-affinity minimum variance~\eqref{eq:bsfano} from \cite{barato15,barato15b}.}
\end{figure}

To explore differences between the constant-force and diffusing-cargo models, we numerically simulate the distribution of the number of net forward steps (hereafter simply the `step number') using the Gillespie algorithm~\cite{gillespie77}, initiated with $\xM = \xC$. For a direct comparison, we choose equal bare rates $\kCF^0=\kM^0$ and equal mean motor velocities,
\begin{equation}
\label{eq:velocity}
v = (k^+ - k^-)\dM \ .
\end{equation}
We rearrange the constant-force model rate constants~\eqref{eq:cfrateconstants} with a mean step number $\mu_n = vt$ to determine the corresponding constant free energy difference
\begin{equation}
\label{eq:offset}
\dGCF = \BUDGET - \ln\left(\frac{\mu_n}{\dM\kCF^0t} + 1\right) \ .
\end{equation}

Figure~\ref{fig:PositionDists} shows the step-number distribution for diffusing-cargo and for constant-force models, with the constant-force model using $\dGCF$ from Eq.~\eqref{eq:offset} and $\mu_n$ from the diffusing-cargo model. The constant-force step-number variance is predicted by motor dynamics using $\sigma_{\text{cf}}^2=2D_{\text{cf}}t$.
Figure~\ref{fig:PositionDists} also shows the Gaussian distribution predicted by Eq.~\eqref{eq:bsfano} with the parameters $\BUDGET=4$ and $N=1$ used by the numerical models, which does not agree with the constant-force step-number distribution. 

Replacing $\BUDGET$ in Eq.~\eqref{eq:bsfano} with a reduced budget $\BUDGET - \dGCF$ does reproduce the constant-force variance $\sigma_{\text{cf}}^2$, demonstrating that the constant-force model agrees with the minimum Fano factor~\eqref{eq:bsfano}. However, the step-number variance of the diffusing-cargo model in Fig.~\ref{fig:PositionDists} is substantially smaller than both the variance of the constant-force model and the minimum variance from naive application of the uncertainty relation~\eqref{eq:bsfano}. The step-number distributions of Fig.~\ref{fig:PositionDists} suggest that the effect of cargo attached to the motor is not well-represented by a constant force and that the cargo allows the motor to circumvent the minimum Fano factor for a constant-affinity cycle~\eqref{eq:bsfano}.

\begin{figure}[tb]
	\centering
	\hspace{-0.0in}
	\begin{tabular}{c}
		\hspace{-0.200in}\includegraphics[width=3.4in]{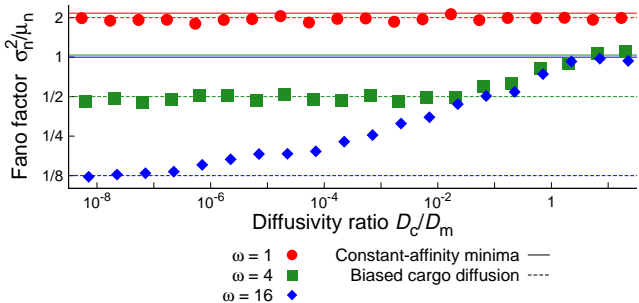}
	\end{tabular}
	\caption{\label{fig:VarianceVsMean}
		{\bf Motor step-number Fano factor,} 
		as function of cargo-motor diffusivity ratio. Points show numerical diffusing-cargo model averaged over $10^3$ samples after a nondimensionalized time in the range $t^{\ast} \equiv \kM^0 t = 10^{-4}-10^9$ chosen to maintain $\mu_{n}\sim100$, thereby providing good statistics and retaining numerical efficiency.
		Colors indicate different free energy budgets $\BUDGET$ per cycle. Other parameters as in Fig.~\ref{fig:PositionDists}. Solid lines show motor-controlled behavior of the predicted minimum Fano factor~\eqref{eq:bsfano} using $N=1$ and $\BUDGET$ as indicated. Dashed curves show cargo-controlled behavior from discrete cargo steps taken to the continuum limit~\eqref{eq:fanocontinuous} with motor step size $\dM$ held fixed. $\Dc/\Dm$ is the ratio of cargo diffusivity $\Dc$ (in the absence of the motor) to cargo-less motor diffusivity $\Dm$. 
	}  
\end{figure}

To investigate the discrepancy between the diffusing-cargo model step-number distribution and Eq.~\eqref{eq:bsfano}, we find the Fano factor for the diffusing-cargo model across several free energy budgets $\BUDGET$ and a wide range of cargo diffusivities (Fig.~\ref{fig:VarianceVsMean}). At low cargo diffusivity, the diffusing-cargo Fano factor is below the constant-affinity minimum Fano factor~\eqref{eq:bsfano}. For the $N=1$ cycle used for the diffusing-cargo model, Eq.~\eqref{eq:bsfano} predicts that as $\BUDGET\to\infty$ the Fano factor approaches unity from above---this prediction is unambiguously not followed by the diffusing-cargo model in Fig.~\ref{fig:VarianceVsMean}. Instead, the diffusing-cargo Fano factor at low cargo diffusivity moves further below the Eq.~\eqref{eq:bsfano} prediction as the free energy budget $\BUDGET$ increases, with the Fano factor for free energy budget $\BUDGET=16$ approximately 1/8 that predicted in Eq.~\eqref{eq:bsfano}.

We investigate low-diffusivity cargo because of the relatively small diffusivities (compared to nanometer-scale motors) of micron-scale vesicles and organelles, which \emph{in vivo} molecular motors tow up to millimeter distances~\cite{hirokawa09}. For cargo with diffusivity much lower than that of the motor, the distribution of motor positions is near the steady state attained for an unmoving cargo,
where
$\langle f_{\text{net}}\rangle = 0$. Chemical driving imposes a forward force $\langle f_{\text{chem}}\rangle = \BUDGET/\dM$ on the motor. To maintain the steady state, the motor is opposed by the force from the elastic motor-cargo linker; averaged over the steady-state distribution of motor positions (conditioned on the fixed cargo position), the linker force is $\langle f_{\text{cargo}}\rangle = -\BUDGET/\dM$. Thus, low-diffusivity cargo experiences an average forward-directed force $\langle f\rangle=\BUDGET/\dM$.

Figures~\ref{fig:PositionDists} and \ref{fig:VarianceVsMean} show diffusing-cargo model Fano factors below the mimimum Fano factors for constant-affinity dynamics~\eqref{eq:bsfano}. Here we determine the Fano factor for a cargo under a constant forward force, treating a discrete step of the cargo itself as a single-stage cycle. If the cargo takes steps of size $\dC$, with bare rate $\kC^{0}$, its forward and reverse rate constants under constant force $f$ are
\begin{equation}
\kC^{+} = \kC^{0} \ \ \text{and} \ \ \kC^{-} = \kC^{0}e^{-f\dC} \ .
\end{equation}
Cargo under constant force has velocity and diffusivity expressions analogous to Eqs.~\eqref{eq:velocity} and \eqref{eq:diffusivity}, respectively; combining these with flat-landscape cargo diffusivity $\Dc = \kC^0\dC^2$ gives the mean and variance for cargo position under constant force,
\begin{subequations}
	\label{eq:biasedcargo}
	\begin{align}
		\mu_{\text{c}} &= \dM\mu_n = \frac{\Dc t}{\dC}\left(1-e^{-f\dC}\right) \ , \\
		\sigma^2_{\text{c}} &= \dM^2\sigma^2_n = \Dc t\left(1+e^{-f\dC}\right) .
	\end{align}
\end{subequations}
The Fano factor is then (with $f=\BUDGET/\dM$)
\begin{equation}
\label{eq:fanodiscrete}
\frac{\sigma^2_{n}}{\mu_{n}} = \frac{\dC}{\dM}\coth\frac{\BUDGET \dC}{2\dM} \ .
\end{equation}
Substituting $\dC = \dM/N$ recovers~\eqref{eq:bsfano}.
For continuous cargo motion, $\dC\to 0$, simplifying~\eqref{eq:fanodiscrete} to
\begin{equation}
\label{eq:fanocontinuous}
\frac{\sigma^2_{n}}{\mu_{n}} = \frac{2}{\BUDGET} \ .
\end{equation}

In Fig.~\ref{fig:VarianceVsMean}, Eq.~\eqref{eq:fanocontinuous} closely matches the numerical Fano factors for the diffusing-cargo model at low cargo diffusivity, demonstrating that treating cargo motion as a driven cycle leads to accurate prediction of the motor variance.
For low cargo diffusivity, slow cargo movement increases the number $N$ of effective stages per motor cycle.
Using the minimum Fano factor of Barato and Seifert~\eqref{eq:bsfano}, the increase in $N$ reduces the variance, but the variance cannot fall below the minimum required by budget $\BUDGET$.

In Fig.~\ref{fig:VarianceVsMean}, increasing cargo diffusivity causes the motor step-number Fano factor to transition from the cargo-controlled description of Eq.~\eqref{eq:fanocontinuous}, to the motor-controlled prediction of Eq.~\eqref{eq:bsfano} as the cargo diffusivity reaches the cargo-less motor diffusivity $\Dm$. Larger budgets $\BUDGET$ cause wider transitions between cargo-controlled and motor-controlled Fano factors, because higher motor velocity breaks the stationary-cargo approximation at lower $\Dc/\Dm$.

\begin{figure}[tb]
	\centering
	\hspace{-0.0in}
	\begin{tabular}{c}
		\hspace{-0.200in}\includegraphics[width=3.4in]{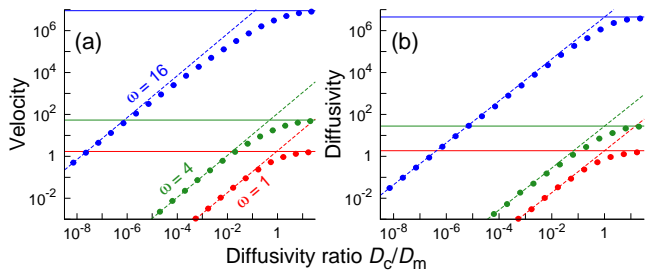}
	\end{tabular}
	\caption{\label{fig:VarianceVsDiffusivity}   
		{\bf Motor velocity and diffusivity are controlled by cargo dynamics when cargo diffusivity is low.} (a) Motor velocity $\mu_n\dM/t^{\ast}$. (b) Motor diffusivity $\sigma^2_n\dM^2/(2t^{\ast})$. Points are from the numerical diffusing-cargo model averaged over $10^3$ samples, with variable cargo diffusivity and $\BUDGET$ as indicated. Other parameters as in Fig.~\ref{fig:VarianceVsMean}. Solid lines show motor-controlled behavior of a motor without cargo, with $\kM^+=e^\BUDGET$ and $\kM^-=1$. Dashed lines show cargo-controlled behavior of biased cargo diffusion~\eqref{eq:biasedcargo}.}
\end{figure}

Figure~\ref{fig:VarianceVsDiffusivity} shows that for $\Dc/\Dm \ll 1$, the motor velocity and diffusivity are predicted by treating the motor-cargo system as cargo pulled by a constant force, while for $\Dc/\Dm \gg 1$, $\mu_{n}$ and $\sigma^2_{n}$ are predicted by treating the system as a cargo-less motor. Although Fig.~\ref{fig:VarianceVsMean} shows that low cargo diffusivity leads to a decreased Fano factor (possibly construed as improved performance), Fig.~\ref{fig:VarianceVsDiffusivity} is a reminder that this cargo-induced precision gain comes at the cost of lower velocity.

We chose a Hookean spring to link motor and cargo. However, our results are robust to variation of several linker properties, such as increased dimensionality of cargo motion, nonzero rest length, finite extensibility, and stiffer spring constant (see Supplementary Material). A previous Langevin model for molecular motors pulling cargo, with force directly determining velocity, finds different behavior in different viscosity regimes~\cite{mckinley12} (analogous to varying cargo diffusivity), further reinforcing that our findings do not depend on model details.

We have used motor rate constants~\eqref{eq:ecrateconstants} for which chemical and spring free energies affect the forward, but not the reverse, transitions~\cite{brown17,brown18}. When instead free energies only affect the motor's reverse transition rates, similar Fano factors are produced (see Supplementary Material).

\section{Discussion}

We have investigated the step-number distribution for simple models of molecular motors carrying out their primary role of towing cargo. We examine both a constant-force model representing cargo in a mean-field manner as reducing the driving force biasing forward motion, and a diffusive-cargo model linking the motor to cargo with its own explicit dynamics.

Although constant-force models of molecular motor operation are widely used~\cite{fisher99,fisher01,thomas01,qian04,lau07,hinczewski13,qian16,wagoner16,hwang17}, they do not accurately reproduce the step-number distribution of diffusive-cargo models, with a model explicitly including diffusive cargo having a substantially lower variance (Fig.~\ref{fig:PositionDists}). The step-number distribution for the constant-force model is consistent with the minimum Fano factor~\eqref{eq:bsfano} derived by Barato and Seifert~\cite{barato15,barato15b}, because the constant-force model satisfies the constant-affinity assumption underpinning Eq.~\eqref{eq:bsfano}. This constant-affinity condition is also fulfilled by the diffusing-cargo model when the cargo dynamics are faster than the motor dynamics, allowing the cargo to quickly relax to near equilibrium following a motor step. However, the diffusive-cargo model produces---for sufficiently large free energy budget $\BUDGET$ and low cargo diffusivity $\Dc$---a significantly lower Fano factor than expected from a naive application of~\eqref{eq:bsfano}
(Fig.~\ref{fig:VarianceVsMean}). Such a low Fano factor occurs because the attachment of diffusive cargo to the motor changes motor dynamics such that the affinity is not constant (\emph{i.e.}\ the ratio of forward to reverse rate constants of the motor depends on the relative locations of motor and cargo), thus violating the constant-affinity assumption in Barato and Seifert~\cite{barato15,barato15}. The attachment of diffusive cargo to a molecular motor can thus increase the precision of forward motor progress.

For low cargo diffusivity, the cargo moves very slowly, and forward progress is well-approximated by assuming the {\em motor} provides a constant forward force on the {\em cargo}. This contrasts with analogous constant-force models that assume the cargo imposes a constant force on the motor. This change in behavior when switching from high to low cargo diffusivity regimes can also be achieved by switching from low to high viscosity while leaving cargo unchanged~\cite{mckinley12}.

Our derived Fano factor for biased cargo diffusion~\eqref{eq:fanocontinuous} is predictive of diffusive-cargo model behavior when cargo diffusivity is low (Fig.~\ref{fig:VarianceVsMean}), and represents a new, more permissive bound on the precision of motor progress. Equation~\eqref{eq:fanocontinuous} eliminates the number of stages in a motor cycle from consideration as a factor limiting motor precision, and is equal to the Barato-Seifert Fano factor~\eqref{eq:bsfano} for $N$ equal to the very large effective number of cargo steps to complete a single motor step~\eqref{eq:fanodiscrete}.
Thus low-diffusivity cargo allows the motor to achieve the minimum Fano factor of a cycle with many stages. The minimum Fano factor of Barato and Seifert~\eqref{eq:bsfano} must be applied carefully, and for a composite system of a motor linked to other elements, the entire system must be considered.

Molecular motor kinetic rates inferred from experiment are consistent with those that improve motor precision in our simulations. For example, Vu \emph{et al.}~\cite{vu16} fit rate constants for a one-stage model of kinesin, suggesting a free energy budget per cycle of $\sim 6k_{\text{B}}T$ (substantially lower than the expected 20$k_{\text{B}}T$ from ATP hydrolysis). This budget is sufficiently large to produce a step-number Fano factor for a motor with \emph{in vivo} cargos ($\Dc/\Dm \approx 10^{-3}$) that significantly differs from that of a motor without cargo (Fig.~\ref{fig:VarianceVsMean}).

Potentially comparable experiments on biomolecular motor systems~\cite{svoboda94,visscher99,cappello03}, although pulling 
cargo ($\sim50\text{nm}$ beads) 
significantly larger than the motor itself, 
only explore diffusivity ratios of $D_{\text{c}}/D_{\text{m}}\sim200$, which for our model is too large to observe increased precision of motor progress (Fig.~\ref{fig:VarianceVsMean}). These experiments are performed in room-temperature water with viscosity $\sim10^{-3}\text{Pa}\cdot\text{s}$; cellular interiors are $10^5-10^6\times$ more viscous~\cite{luby-phelps00,caragine18}, sufficiently suppressing the diffusivity of typical cellular cargo 
such
that we would predict increased motor precision. A comparison of step-number Fano factor between a cargo-less motor and an identical motor with low-diffusivity cargo, both in high-viscosity conditions, would distinguish intrinsic motor behavior from the cargo-induced Fano factor decrease our work predicts, but we are unaware of any existing studies with such data.

The constant-force model assumes that the free energy difference $\Delta G_{\text{cf}}$ modeling the effect of cargo is diverted to moving the motor against the constant force, and is unavailable to bias motor dynamics in the forward direction. For a motor pulling diffusive cargo, free energy is initially stored in the elastic linker connecting motor and cargo, but is ultimately dissipated as the cargo relaxes, and would not generally be available for reversible extraction at a later time~\cite{wang02b}. The assumption that the free energy is stored, when it is instead dissipated, leads to an underestimate of the free energy dissipation budget per motor cycle, which is remedied by explicit consideration of cargo. The elastic restoring force and energy storage of the linker play an important role for a reduced cargo-controlled Fano factor, as we demonstrate with `deterministic' cargo dynamics (an unphysical cargo with drag but not stochastic diffusion) in the Supplementary Material. 

In vivo motors often work in teams of two or more to pull cargo through the cell~\cite{mallik13}. Following Eqs.~\ref{eq:biasedcargo}-\ref{eq:fanocontinuous} with two motors, we expect a halved Fano factor of $\sigma_{n}^2/\mu_{n} = 1/\BUDGET$, consistent with previous work exploring multiple motors~\cite{mckinley12}, suggesting multiple motors would further increase precision of motor progress.

\begin{acknowledgments}
The authors thank Miranda Louwerse (SFU Chemistry), Steve Large, John Bechhoefer, and especially Nancy Forde (SFU Physics) for useful discussions and feedback. This work was supported by a Natural Sciences and Engineering Research Council of Canada (NSERC) Discovery Grant, by funds provided by the Faculty of Science, Simon Fraser University through the President's Research Start-up Grant, by a Tier II Canada Research Chair, and by WestGrid (www.westgrid.ca) and Compute Canada Calcul Canada (www.computecanada.ca).
\end{acknowledgments}

\appendix
\section{Overdamped cargo dynamics under force}
We describe cargo dynamics using rate constants that obey detailed balance and produce expected macroscopic overdamped response to applied forces.

The forward ($+$) and reverse ($-$) cargo transition rate constants are
\begin{equation}
\label{eq:cargorateconstants}
\kC^{\pm}=
\begin{cases}
\kC^{0}, &\text{if }\Delta G_{\text{sc}}^{\pm}(\xM-\xC) \leq 0\\
\kC^{0}e^{-\Delta G_{\text{sc}}^{\pm}(\xM-\xC)}, &\text{if }\Delta G_{\text{sc}}^{\pm}(\xM-\xC) > 0 \ ,
\end{cases}
\end{equation}
where $x_{\text{m}}$ is the motor position, $x_{\text{c}}$ is the cargo position, and $\kC^{0}$ is the bare rate constant for the cargo. $\Delta G_{\text{sc}}^{\pm} = \mp k\dC(\xM - \xC) + \tfrac{1}{2}k\dC^2$ is the change in spring free energy $G_{\text{s}}$ due to a forward or reverse cargo step of size $\dC$.

Such dynamics produce a mean velocity of the cargo:
\begin{equation}
v = (k_{\text{c}}^+ - k_{\text{c}}^-)d_{\text{c}} \ .
\end{equation}
Substituting Eq.~\ref{eq:cargorateconstants}, when the linker is stretched beyond its rest length (and hence $\Delta G_{\text{sc}}^- > 0$ and $\Delta G_{\text{sc}}^+ < 0$), 
\begin{equation}
v = k_{\text{c}}^0 d_{\text{c}}\left(1 - e^{-\Delta G^-_{\text{sc}}}\right) \ .
\end{equation}
For $\Delta G^-_{\text{sc}} \ll 1$ (due to small steps $d_{\text{c}}$),
\begin{equation}
v \simeq k_{\text{c}}^0 d_{\text{c}} \Delta G^-_{\text{sc}} \ .
\end{equation}
The change in spring free energy over a step of size $d_{\text{c}}$ is $\Delta G^-_{\text{sc}} \simeq k d_{\text{c}}(x_{\text{m}} - x_{\text{c}})$ (neglecting higher order terms in $d_{\text{c}}$), and the cargo diffusivity on a flat energy landscape is $D_{\text{c}} = k_{\text{c}}^0 d_{\text{c}}^2$, thus
\begin{equation}
v \simeq D_{\text{c}}k(x_{\text{m}} - x_{\text{c}}) \ .
\end{equation}
$D = 1/\zeta$ for friction coefficient $\zeta$ (and we work in units of $k_{\text{B}}T$), so
\begin{equation}
v \simeq \frac{k}{\zeta}(x_{\text{m}} - x_{\text{c}}) \ .
\end{equation}
This expression---of cargo mean velocity when coupled to the motor via an elastic spring---is identical to previous descriptions of overdamped response to an elastic force~\cite{mazonka99}.

The main text numerical results link motor and cargo via a spring whose spring constant satisfies $kd_{\text{m}}^2=1$. This corresponds to spring constant
\begin{equation}
k = \frac{k_{\text{B}}T}{(8\text{nm})^2} = \frac{4.1\text{pN}\cdot\text{nm}}{64\text{nm}^2} = 0.064\text{pN}/\text{nm} \ .
\end{equation}

\section{Molecular motor diffusivity}
Vu \emph{et al}~\cite{vu16} find forward and reverse stepping rate constants for kinesin of $k^+\simeq133\text{s}^{-1}$ and $k^-\simeq0.4\text{s}^{-1}$, respectively, giving a motor diffusivity
\begin{subequations}
	\begin{align}
		\Dm &= \frac{\dM^2}{2}\left(k^+ + k^-\right)\\
		&= \frac{(0.008\mu\text{m})^2}{2}\left(133\text{s}^{-1} + 0.4\text{s}^{-1}\right)\\
		&= 4.27\times10^{-3}\mu\text{m}^2/\text{s} \ .
	\end{align}
\end{subequations}

\section{Three-dimensional cargo motion and linker with nonzero rest length}

\begin{figure}[tbp] 
	\centering
	\hspace{-0.0in}
	\begin{tabular}{c}
		\hspace{-0.100in}\includegraphics[width=3.5in]{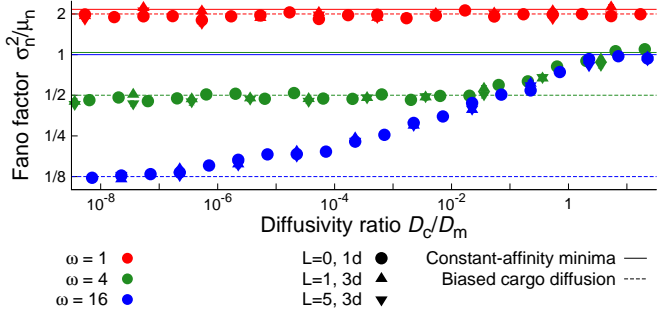}
	\end{tabular}
	\caption{\label{fig:MoreReal}
		{\bf Motor precision is essentially insensitive to dimensionality of cargo motion and linker rest length.}
		Motor step-number Fano factor, as a function of cargo-motor diffusivity ratio, for forward-labile rate constants. Points show numerical diffusing-cargo model averaged over $10^3$ samples after a time chosen to maintain a mean number of motor steps $\mu_n \sim 100$, thereby providing good statistics and retaining numerical efficiency. Colors indicate different free energy budgets $\BUDGET$ per cycle. Circles: one-dimensional cargo motion and linker with zero rest length. Up triangles: three-dimensional cargo motion and linker with a rest length of $d_{\text{m}}$. Down triangles: three-dimensional cargo motion and linker with a rest length of $5d_{\text{m}}$. Solid lines show motor-controlled behavior of the predicted minimum Fano factor~(5)
		and dashed curves show cargo-controlled behavior from discrete-step cargo dynamics taken to the continuum limit~(11).
		Other parameters as in Fig.~2.
	}  
\end{figure}

In all theoretical and numerical results in the main text, cargo motion is constrained to one dimension (the direction of motor motion, $x$), coupled to the motor via a linker with zero rest length. Figure~\ref{fig:MoreReal} analyzes how the Fano factor changes when cargo diffuses in three dimensions, with a linker with nonzero rest length. The cargo dynamics in the $y$ and $z$ directions are constructed to also obey Eq.~(4).
The free energy of the spring linking motor and cargo is $G_{\text{s}} = \tfrac{1}{2}k(x_{\text{m}} - x_{\text{c}} - L)^2$, where $L$ is the linker rest length. In Fig.~\ref{fig:MoreReal}, three-dimensional cargo motion and nonzero linker rest length produce motor precision indistinguishable from that for one-dimensional cargo motion and zero linker rest length.

\section{Worm-like chain linker}

\begin{figure}[tbp] 
	\centering
	\hspace{-0.0in}
	\begin{tabular}{c}
		\hspace{-0.100in}\includegraphics[width=3.5in]{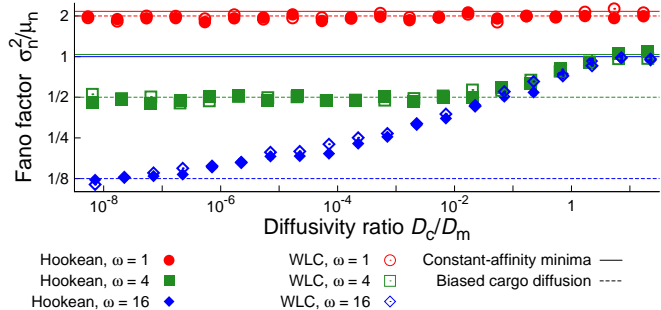} 
	\end{tabular}
	\caption{\label{fig:wlc}
		{\bf Motor precision is insensitive to finite extensibility of cargo-motor linker.} 
		Motor step-number Fano factor, as a function of cargo-motor diffusivity ratio, for forward-labile rate constants. Points show numerical diffusing-cargo model averaged over $10^3$ samples after a time chosen to maintain $\mu_n\sim100$, thereby providing good statistics and retaining numerical efficiency. Colors indicate different free energy budgets $\BUDGET$ per cycle. Filled points show Hookean linker with $k^*\equiv kd_{\text{m}}^2 = 1$. Open points are for worm-like chain linker, with linker energy in Eq.~\ref{eq:wlcenergy} using persistence length $\ell_{\text{p}} = 0.1d_{\text{m}}$ and contour length $\ell_{\text{c}}=10d_{\text{m}}$. Solid lines show motor-controlled behavior of the predicted minimum Fano factor~(5)
		and dashed curves show cargo-controlled behavior from discrete-step cargo dynamics taken to the continuum limit~(11).
		Other parameters as in Fig.~2.
	}  
\end{figure}

In all theoretical and numerical results in the main text, the linker between motor and cargo is a Hookean spring. However, linkers between biomolecular motors and their cargo are not infinitely extensible and thought to deviate from a simple Hookean force response~\cite{mckinley12,palacci16}. We explore the motor behavior with a finite-length linker, modeled as a worm-like chain, with free energy 
\begin{equation}
\label{eq:wlcenergy}
G_{\text{s}} = \int_0^{x_{\text{m}} - x_{\text{c}}}f_{\text{WLC}}(x)\mathrm{d}x \ ,
\end{equation}
where
\begin{equation}
f_{\text{WLC}}(x) = \frac{1}{\ell_{\text{p}}}\left[\frac{1}{4}\left(1 - \frac{x}{\ell_{\text{c}}}\right)^{-2} - \frac{1}{4} + \frac{x}{\ell_{\text{c}}}\right]
\end{equation}
is the interpolated worm-like chain force for end separation $x$, persistence length $\ell_{\text{p}}$, and contour length $\ell_{\text{c}}$. We use $\ell_{\text{c}}/d_{\text{m}} = 10$ to match kinesin linker length of approximately 78nm~\cite{jeney04} (kinesin step size is $d_{\text{m}}=8$nm) and $\ell_{\text{p}}/d_{\text{m}} = 0.1$, such that at small extensions the linker is approximately a Hookean spring with $k=1$, for direct comparison with results in the main text. In Fig.~\ref{fig:wlc}, worm-like chain and Hookean spring linkers lead to extremely similar motor precisions. 

\section{Linker spring constant}

\begin{figure}[tbp] 
	\centering
	\hspace{-0.0in}
	\begin{tabular}{c}
		\hspace{-0.100in}\includegraphics[width=3.5in]{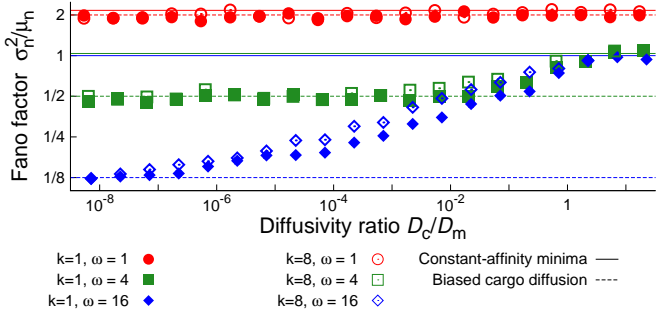}
	\end{tabular}
	\caption{\label{fig:k8}
		{\bf Variation of linker spring constant makes little difference in motor precision.} Motor step-number Fano factor, as a function of cargo-motor diffusivity ratio, for forward-labile rate constants. Points show numerical diffusing-cargo model averaged over $10^3$ samples after a time chosen to maintain $\mu_n\sim100$, thereby providing good statistics and retaining numerical efficiency. Colors indicate different free energy budgets $\BUDGET$ per cycle. Filled points have linker spring constant $k^*\equiv kd_{\text{m}}^2 = 1$, and open points have $kd_{\text{m}}^2 = 8$. Solid lines show motor-controlled behavior of the predicted minimum Fano factor~(5)
		and dashed curves show cargo-controlled behavior from discrete-step cargo dynamics taken to the continuum limit~(11).
		Other parameters as in Fig.~2.
	}  
\end{figure}

In our main results, the linker between motor and cargo is a Hookean spring with spring constant satisfying $kd_{\text{m}}^2=1$, which for kinesin, with step size $d_{\text{m}} = 8\text{nm}$, corresponds to $k=0.064$pN/nm. The experimental literature reports spring constants for the kinesin linker ranging from 0.32pN/nm~\cite{bergman18} to 0.5pN/nm~\cite{li12}. Therefore, we also explore $kd_{\text{m}}^2=8$, corresponding to $k=0.512$pN/nm. Figure~\ref{fig:k8} shows that $kd_{\text{m}}^2=1$ and $kd_{\text{m}}^2=8$ lead to similar motor precisions.

\section{Reverse-labile motor dynamics}
Free energy dissipation $\BUDGET$ drives motor dynamics by increasing the ratio between forward ($k^+$) and reverse ($k^-$) rate constants according to the generalized detailed balance condition, $\BUDGET = \ln(k^+/k^-)$. However, generalized detailed balance leaves unspecified whether dissipation increases $k^+$, decreases $k^-$, or some combination of the two. The main text uses forward-labile~\cite{brown17,brown18} (FL) rate constants~(3),
the extreme where dissipation only increases forward rate constants, leaving reverse rate constants unchanged. Here we explore reverse-labile (RL) rate constants,
\begin{equation}
\label{eq:reverselabilerateconstants}
\kM^{+} = \kM^{0} \ \ \text{and} \ \ \kM^{-} = \kM^{0}e^{-\BUDGET - \Delta G_{\text{sm}}^-(\xM,\xC)} \ ,
\end{equation}
the opposite extreme where dissipation only decreases reverse rate constants, leaving forward rate constants unchanged. We use $k^0_{\text{m,RL}} = e^{\BUDGET}k^0_{\text{m,FL}}$ such that forward-labile and reverse-labile rate constants are equal in the absence of cargo.

\begin{figure}[tbp] 
	\centering
	\hspace{-0.0in}
	\begin{tabular}{c}
		\hspace{-0.100in}\includegraphics[width=3.5in]{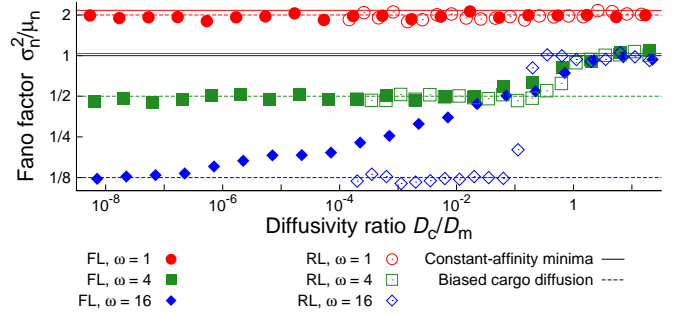}
	\end{tabular}
	\caption{\label{fig:ReverseLabile}
		{\bf Reverse-labile motors and forward-labile motors both transition from cargo- to motor-controlled Fano factor,} 
		as a function of cargo-motor diffusivity ratio. Points show numerical diffusing-cargo model averaged over $10^3$ samples after a time chosen to maintain $\mu_n\sim100$, thereby providing good statistics and retaining numerical efficiency. Filled points show forward-labile kinetics~(3),
		and open points show reverse-labile kinetics~\eqref{eq:reverselabilerateconstants}. Colors indicate different free energy budgets $\BUDGET$ per cycle. Solid lines show motor-controlled behavior of the predicted minimum Fano factor~(5)
		and dashed curves show cargo-controlled behavior from discrete-step cargo dynamics taken to the continuum limit~(11).
		Other parameters as in Fig.~2.
	}  
\end{figure}

Figure~\ref{fig:ReverseLabile} compares motors with forward- and reverse-labile rate constants, showing that both models transition from a low cargo-controlled Fano factor~(11)
at low cargo diffusivity to a high motor-controlled Fano factor~(5)
at high cargo diffusivity. 
One distinction is that reverse-labile motors with $\BUDGET=16$ and $\BUDGET=4$ transition between the limiting Fano factors over a narrower range of cargo diffusivity.
This narrower transition 
is rationalizable in terms of the different sensitivities of FL and RL motors to applied resisting force.

\begin{figure}[tbp] 
	\centering
	\hspace{-0.0in}
	\begin{tabular}{c}
		\hspace{-0.100in}\includegraphics[width=3.5in]{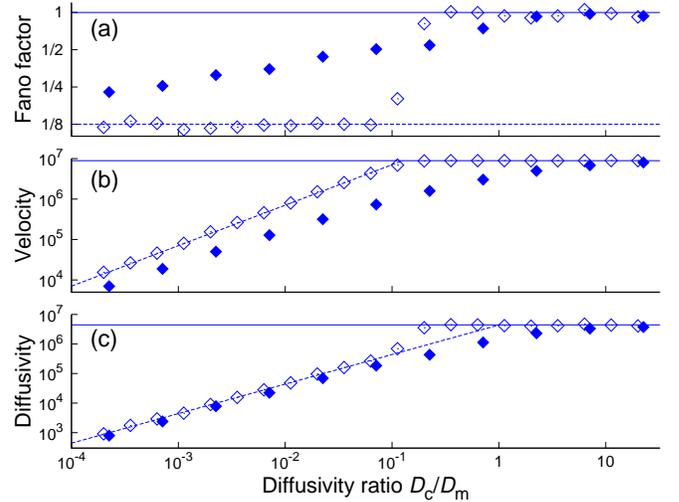}
	\end{tabular}
	\caption{\label{fig:ReverseLabileVariances}
		{\bf 
			Velocity and diffusivity transition more sharply in reverse-labile motors than in forward-labile motors.}
		(a) Motor Fano factor $\sigma_{\text{n}}^2/\mu_{\text{n}}$, (b) motor velocity $\mu_n\dM/t^{\ast}$, and (c) motor diffusivity $\sigma^2_n\dM^2/(2t^{\ast})$. Solid line shows motor-controlled behavior and dashed line cargo-controlled behavior. Points are from the numerical diffusing-cargo model averaged over $10^3$ samples, with variable cargo-motor diffusivity ratio and $\BUDGET$ as indicated. Filled points show forward-labile kinetics~(3)
		and open points show reverse-labile kinetics~\eqref{eq:reverselabilerateconstants}. All lines and points are for $\BUDGET=16$ per cycle. Other parameters as in Fig.~2.
	}  
\end{figure}

Figure~\ref{fig:ReverseLabileVariances}b shows that RL motor velocity follows the cargo-controlled velocity until the motor-controlled velocity is reached, while FL motor velocity deviates below the cargo-controlled velocity as it approaches the motor-controlled velocity. Figure~\ref{fig:ReverseLabileVariances}c shows that when the RL motor reaches the motor-controlled velocity, the motor diffusivity rapidly switches to the motor-controlled diffusivity. Together, these lead to a narrower transition for the RL step-number Fano factor than the FL step-number Fano factor.

Why does RL stick closely to the cargo-controlled velocity in Fig.~\ref{fig:ReverseLabileVariances}b, while FL deviates below? 
FL and RL motors with identical cargo-less diffusivity $D_{\rm m}$ have bare rate constants satisfying $k^0_{\text{m,RL}} = e^{\BUDGET}k^0_{\text{m,FL}}$, producing equal motor-controlled velocities in the absence of cargo. When linked to cargo and in the cargo-controlled regime, the motor predominantly occupies states that balance the motor free energy budget with linker energy, $\BUDGET \simeq \Delta G_{\text{sm}}$. In these dominant states, an RL motor has rate constants a factor $e^{\BUDGET}$ greater than the equivalent FL motor with identical cargo-less $D_{\rm m}$. With much higher rate constants, RL more quickly explores the steady-state distribution of motor positions, and is able to populate this steady-state distribution at higher $D_{\text{c}}/D_{\text{m}}$ than FL.

When in the cargo-controlled regime, the cargo lags the motor by a distance $\sim \BUDGET/(k d_{\text{m}})$. Once the motor-controlled velocity is reached, further increases in $D_{\text{c}}/D_{\text{m}}$ decrease this lag. With this decrease in cargo distance behind the motor, the reverse rate constant of the RL motor substantially decreases, leaving the motor dynamics dominated by a much larger forward rate constant, and leading to a relatively sudden jump in the motor diffusivity in Fig.~\ref{fig:ReverseLabileVariances}c.

\section{Deterministic cargo dynamics}

To probe how the diffusive nature of motor cargo impacts motor precision, we explored a model of deterministic cargo dynamics. Rather than Eq.~(4),
we 
now describe the dynamics with
$\mathrm{d} x_{\text{c}}/\mathrm{d} t = f/\zeta_{\text{c}}$, for linker force $f$ on the cargo and friction coefficient $\zeta_{\text{c}}$. 
We implement this deterministic cargo motion by numerically incrementing time some small increment $\Delta t$ and moving cargo according to $x_{\text{c}}(t+\Delta t) = x_{\text{c}}(t) + f(t)\Delta t/\zeta_{\text{c}}$.
For instantaneous motor forward rate constant $k_{\text{m}}^+(t)$ and reverse rate constant $k_{\text{m}}^-(t)$, over time increment $\Delta t$ the motor takes a step with probability $[k_{\text{m}}^+(t) + k_{\text{m}}^-(t)]\Delta t$. 
The time increment $\Delta t$ is chosen to be the smallest of: the time increment over which the cargo moves a distance $0.001d_{\text{m}}$, the time increment such that $k_{\text{m}}^+(t)\Delta t=0.001$, and the time increment such that $k_{\text{m}}^-(t)\Delta t=0.001$.

\begin{figure}[tbp] 
	\centering
	\hspace{-0.0in}
	\begin{tabular}{c}
		\hspace{-0.100in}\includegraphics[width=3.5in]{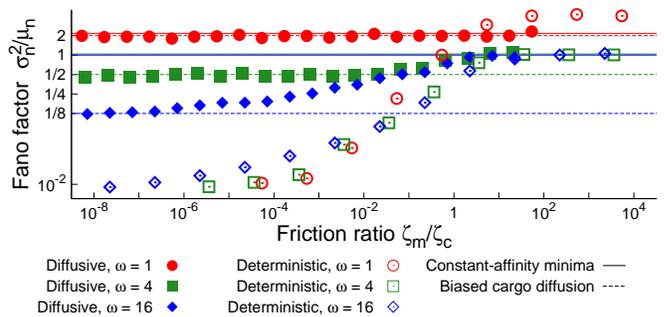}
	\end{tabular}
	\caption{\label{fig:deterministic}
		{\bf Motors with deterministic cargo have higher precision compared to diffusive cargo.}
		Motor step-number Fano factor, as a function of motor-cargo friction ratio $\zeta_{\text{c}}/\zeta_{\text{m}}$, for forward-labile rate constants. Both diffusing-cargo and deterministic cargo models are averaged over $10^3$ samples after a nondimensionalized time $t^{\ast} \equiv k_{\text{m}}^0 t$ chosen to maintain mean motor step-number $\mu_n\sim100$
		Colors indicate different free energy budgets $\BUDGET$ per cycle. Filled points show diffusive cargo dynamics and open points show deterministic cargo dynamics. 
		For motor and diffusive cargo, friction $\zeta$ and diffusivity $D$ are related by the Einstein relation, $\zeta = 1/D$.
		Solid lines show motor-controlled behavior of the predicted minimum Fano factor~(4)
		and dashed curves show cargo-controlled behavior from discrete-step cargo dynamics taken to the continuum limit~(11).
		Other parameters as in Fig.~2.
	}  
\end{figure}

Figure~\ref{fig:deterministic} compares deterministic cargo dynamics to diffusive cargo dynamics. For high-friction cargo, motor step-number precision is much higher (\emph{i.e.}, Fano factor is much lower) for deterministic dynamics -- this is expected, as random cargo motion no longer contributes to the position dispersion of the motor-cargo system. For low-friction cargo, the high motor free energy budgets ($\BUDGET=4$ and $\BUDGET=16$) reach the Barato-Seifert limit we expect.

Anomalously, for low-friction cargo, a motor with $\BUDGET=1$ coupled to a deterministic cargo has a higher Fano factor than when coupled to a diffusive cargo. 
This phenomenon is rationalizable in terms of the model, but ultimately is unphysical, because for these motor-cargo friction ratios the model simultaneously assumes that the cargo instantly responds to motor steps, yet also that the motor takes an entire step before the cargo can respond. 

For this low-friction deterministic cargo dynamics, the cargo nearly instantaneously reaches the motor position following a motor step, and the cargo remains at the motor position until the motor takes another step. In Eq.~(3),
the forward motor stepping rate constant is
\begin{equation}
k_{\text{m}}^+ = k_{\text{m}}^0e^{\BUDGET - \Delta G_{\text{sm}}^+(x_{\text{m}} - x_{\text{c}})} \ ,
\end{equation}
with change in linker free energy
\begin{equation}
\Delta G_{\text{sm}}^{+}(\xM - \xC) = k\dM(\xM - \xC) + \tfrac{1}{2}k\dM^2
\end{equation} 
over a motor step. 
With $x_{\text{m}}-x_{\text{c}}=0$ and $kd_{\text{m}}^2=1$,  
$\Delta G_{\text{sm}}^+ = 1/2$, such that $k_{\text{m}}^+ = k_{\text{m}}^0 e^{\BUDGET-\frac{1}{2}}$. The mean step number is
\begin{equation}
\mu_n = (k^+_{\text{m}} - k^-_{\text{m}})t \ ,
\end{equation}
and the step-number variance is
\begin{equation}
\sigma_n^2 = (k_{\text{m}}^+ + k_{\text{m}}^-) t \ .
\end{equation}
The step-number Fano factor for deterministic low-friction cargo is thus
\begin{align}
	\frac{\sigma_{\text{n}}^2}{\mu_n} &= \frac{k_{\text{m}}^+ + k_{\text{m}}^-}{k_{\text{m}}^+ - k_{\text{m}}^-} = \frac{e^{\BUDGET - \tfrac{1}{2}} + 1}{e^{\BUDGET - \tfrac{1}{2}} - 1}\\
	&=  \coth\frac{\BUDGET - \tfrac{1}{2}}{2} \ .
\end{align}
For $\BUDGET=4$ and $\BUDGET=16$, this deterministic low-friction-cargo Fano factor is approximately 1, which is indistinguishable from $\coth(\BUDGET/2)\simeq 1$ for diffusive cargo (Eq.~(5)
in main text). However for $\BUDGET=1$, the deterministic low-friction-cargo Fano factor, $\coth[(\BUDGET-\tfrac{1}{2})/2]\simeq 4$, while the diffusive-cargo Fano factor is $\coth(\BUDGET/2)\simeq 2$, so that for low-friction cargo and low budgets $\BUDGET$ the deterministic-cargo step-number Fano factor is larger than the diffusive-cargo step-number Fano factor.

Ultimately, such a model of deterministic cargo dynamics is unphysical: stochastic fluctuations of the surrounding medium are the source of both the frictional drag on the cargo and its stochastic fluctuations, so it is unphysical to posit one without the other. Nevertheless, the model helps disentangle the distinct effects on motor precision of slow cargo relaxation and cargo fluctuations. 


%

\end{document}